\documentclass[10pt, conference, compsocconf]{IEEEtran}

\usepackage{textcomp}
\usepackage{listings}
\usepackage{color}

\ifCLASSOPTIONcompsoc
\usepackage[caption=false,font=footnotesize,labelfont=sf,textfont=sf]{subfig}
\else
\usepackage[caption=false,font=footnotesize]{subfig}
\fi

\definecolor{codegreen}{rgb}{0,0.6,0}
\definecolor{codegray}{rgb}{0.5,0.5,0.5}
\definecolor{codepurple}{rgb}{0.58,0,0.82}
\definecolor{backcolour}{rgb}{0.95,0.95,0.92}
 
\lstdefinestyle{mystyle}{
    backgroundcolor=\color{backcolour},   
    commentstyle=\color{codegreen},
    keywordstyle=\color{magenta},
    numberstyle=\tiny\color{codegray},
    stringstyle=\color{codepurple},
    basicstyle=\footnotesize,
    breakatwhitespace=false,         
    breaklines=true,                 
    captionpos=b,                    
    keepspaces=true,                 
    numbers=left,                    
    numbersep=5pt,                  
    showspaces=false,                
    showstringspaces=false,
    showtabs=false,                  
    upquote=true,
    tabsize=2
}
 
\ifCLASSINFOpdf
  \usepackage[pdftex]{graphicx}
  \DeclareGraphicsExtensions{.pdf,.jpeg,.png}
\else
\fi

\begin{document}

\title{libhclooc: Software Library Facilitating Out-of-core Implementations of Accelerator Kernels on Hybrid Computing Platforms}

\author{\IEEEauthorblockN{Daniel Hanlon\IEEEauthorrefmark{1},
Hamidreza Khalighzadeh\IEEEauthorrefmark{1}, 
Ravi Reddy Manumachu\IEEEauthorrefmark{2},
and
Alexey Lastovetsky\IEEEauthorrefmark{2}}
\IEEEauthorblockA{School of Computer Science\\
University College Dublin\\
Dublin, Ireland\\
Email: \IEEEauthorrefmark{1}\{daniel.hanlon,hamidreza.khalighzadeh\}@ucdconnect.ie,
\IEEEauthorrefmark{2}\{ravi.manumachu,alexey.lastovetsky\}@ucd.ie}}

\maketitle

\begin{abstract}
Hardware accelerators such as Graphics Processing Units (GPUs), Intel Xeon Phi co-processors (PHIs), and Field-Programmable Gate Arrays (FPGAs) are now ubiquitous in extreme-scale high performance computing (HPC), cloud, and Big data platforms to facilitate execution of workloads that demand high energy efficiency. They present unique interfaces and programming models therefore posing several limitations, which must be addressed to facilitate execution of large workloads. There is no library providing a unifying interface that allows programmers to write reusable out-of-core implementations of their data-parallel kernels that can run efficiently on different mainstream accelerators such as GPUs, PHIs, and FPGAs. We address this shortage in this paper.

We present a library called \emph{libhclooc}, which provides a unifying interface facilitating out-of-core implementations for data-parallel kernels on the three different mainstream accelerators (GPUs, Intel Xeon Phis, FPGAs). We implement out-of-core matrix-matrix multiplication (MMOOC) using the libhclooc API and demonstrate its superior performance over vendor implementations. We show that it suffers from a maximum overhead of 10\%, 4\%, and 8\% (due to abstraction) compared to the state-of-the-art optimized implementations for Nvidia K40c GPU, Nvidia P100 PCIe GPU, and Intel Xeon Phi 3120P respectively. We also show that using libhclooc API reduces the number of lines of code (LOC) by 75\% thereby drastically improving programmer productivity.
\end{abstract}


\IEEEpeerreviewmaketitle

\section{Introduction}

Extreme-scale high performance computing (HPC), cloud, and Big data platforms today feature hybrid nodes containing multicore CPU processors and one or more accelerators such as Graphics Processing Units (GPUs), Intel Xeon Phi co-processors (PHIs), and Field-Programmable Gate Arrays (FPGAs) to facilitate execution of workloads that demand high energy efficiency (high performance and low energy consumption).

There are several challenges posed to execution of large problem sizes on these hybrid nodes arising from the tight integration of the accelerators with multicore CPUs via PCI-E communication links and disparate streaming interfaces. The challenges are summarized below:
\begin{itemize}
\item \textbf{Limited main memory of accelerators.} The size of main memory in accelerators is small compared to that of the host multicore CPU connected to it. For example, consider the Top500 list of supercomputers \cite{Top500}. The Tianhe-2 supercomputer ranked two is composed of Intel Ivybridge multicore CPUs, which support 768 GB per socket, and Intel Xeon Phi 31S1P accelerator, which provides only 8 GB main memory. The TSUBAME3.0 ranked thirteen comprises of Intel Broadwell CPUs, which can support 1.54TB main memory, and NVIDIA Tesla P100 SXM2 accelerators, which support only 16GB of main memory. Therefore, to execute large data-parallel workloads using these accelerators, out-of-core (or out-of-card) implementations are necessary.

\item \textbf{Communication and computation overlap.} Accelerators such as GPUs provide advanced hardware support to facilitate overlap of data transfers between host and device and computations on the device. For example, modern Nvidia GPUs (K40, K80, P100, etc) provide three engines: two copy engines, one for host-to-device transfers and another for device-to-host transfers, and a kernel engine. Therefore, libraries aiming to provide efficient out-of-core implementations must take into account the differences in hardware support for effective communication-computation overlap to optimize their software pipelines for out-of-core implementations.

\item \textbf{Disparate streaming interfaces.} Different vendors provide different streaming interfaces for data transfers between host CPU and accelerators and kernel executions on the accelerators. Nvidia provides CUDA streams and events \cite{cuda-streams} for its GPUs; Intel provides offload streams \cite{offload-streams} for Intel Xeon Phis, which will be replaced soon by hStreams \cite{hstreams}; OpenCL command queues \cite{opencl-commandqueues} are typically used for FPGAs. The interfaces exhibit significant differences in their APIs, programming models, and memory management. For example., Intel offload streams offer a combination of API functions and compiler pragmas. This wide disparity in the interfaces means that the programmers need to write different implementations for their out-of-core accelerator kernels with little code reuse. This can drastically impact programmer's productivity.
\end{itemize}

There is severe shortage of software libraries that address these challenges comprehensively. The CUBLAS-XT library \cite{CUBLAS-XT} provides a set of BLAS routines that support out-of-core operation. MAGMA \cite{MAGMA} provides out-of-core dense LU, Cholesky, and QR factorizations. Victream \cite{Suzuki2017} is a directed acyclic graph (DAG) computing framework for out-of-core computations on multiple GPUs.

There is no library providing a unifying interface that allows programmers to write reusable out-of-core implementations of their data-parallel kernels that can run efficiently on different mainstream accelerators such as GPUs, PHIs, and FPGAs. We address this shortage in this paper.

We present a library called \emph{libhclooc}, which provides a unifying interface facilitating out-of-core implementations for data-parallel kernels on the three different mainstream accelerators (GPUs, Intel Xeon Phis, FPGAs). Its fundamental building blocks are CUDA streams and events that allow concurrent utilization of the copy and execution engines provided in NVidia GPUs \cite{cuda-streams}, Intel offload streams \cite{offload-streams} for Intel Xeon Phis, and OpenCL command queues \cite{opencl-commandqueues} for FPGAs. 

We implement out-of-core matrix-matrix multiplication (MMOOC) using the libhclooc API and demonstrate that it outperforms vendor implementations (CUBLAS-XT \cite{CUBLAS-XT}) and suffers from a maximum overhead of 10\%, 4\%, and 8\% (due to abstraction) compared to the state-of-the-art accelerator-specific optimized implementations for Nvidia K40c GPU, Nvidia P100 PCIe GPU, and Intel Xeon Phi 3120P respectively. We show that using libhclooc API reduces the number of lines of code (LOC) by 75\% thereby drastically improving programmer productivity.

To summarize, our main contributions in this paper are:
\begin{itemize}
\item Software library \emph{libhclooc} that presents a unifying interface for CUDA streams, Intel offload streams, and OpenCL command queues thereby allowing programmers to write efficient reusable out-of-core implementations of their data-parallel kernels for three mainstream accelerators (GPUs, Intel Xeon Phis, FPGAs).
\item Implementation of an out-of-core matrix-matrix multiplication using libhclooc that executes on three different accelerators with an abstraction overhead in the library of 10\%. The implementation reduces the number of lines of code (LOC) by 75\%.
\item Implementation of an out-of-core matrix-matrix multiplication using libhclooc that outperforms the CUBLAS-XT implementation on Nvidia P100 PCIe GPU by 4x.
\end{itemize}

The paper is organized as follows. Section \ref{relatedwork} presents related work. Section \ref{libhcloocoverview} contains the overview of interfaces of libhclooc. Section \ref{libhcloodesign} presents design and implementation details of libhclooc. Section \ref{libhcloocmxm} describes MMOOC using libhclooc API. Section \ref{experimentalresults} contains the experimental results. Section \ref{conclusions} concludes the paper.

\section{Related Work} \label{relatedwork}

We organize our literature survey into three categories. First category contains research works proposing out-of-core techniques and implementations for accelerator kernels. Second category reviews libraries specifically for out-of-core implementations of accelerator kernels. Final category reviews research proposals for streams supporting execution of scientific kernels using multicore CPUs and accelerators. 

\subsection{Out-of-core Techniques and Implementations for Accelerator Kernels}

Gu et al. \cite{Gu2011} present an out-of-core implementation of FFT kernel for a single GPU where they overlap kernel computations on the GPU and communications over the PCI-E bus. Mu et al. \cite{mu2014higher} propose an out-of-core algorithm for LU decomposition.

Ziming et al. \cite{Ziming2012},\cite{zhong2014optimization} propose an out-of-core implementation for matrix multiplication routine (DGEMM) for NVidia GPU. Wu et al. \cite{Wu2016} presented an out-of-core dense matrix multiplication implementation for CPU-GPU platforms similar to \cite{Ziming2012},\cite{zhong2014optimization}.

Sabne et al. \cite{Sabne2013} present a computation splitting technique that automatically adjusts the number of pipeline stages to improve the performance of out-of-core implementations on multiple GPUs attached to the same host CPU.

Shirahata et al. \cite{Shirahata2014} present out-of-core techniques for large-scale graph processing applications for heterogeneous GPU-based clusters.

Kabir et al. \cite{Kabir2017}, Haidar et al. \cite{Haidar2017} propose out-of-core implementations for large dense singular value decompositions (SVD) for CPU architectures.

Yamazaki et al. \cite{Yamazaki2017} present out-of-core algorithms to factorize a symmetric indefinite matrix for CPU and GPU architectures.

Hamid et al. \cite{Khaleghzadeh2018} present out-of-core implementations for matrix multiplication for three accelerators (GPU, PHI, FPGA). However, the implementations suffer from heavy code duplication and use disparate interfaces such as CUDA streams \cite{cuda-streams}, PHI streams \cite{offload-streams}, and OpenCL command queues \cite{opencl-commandqueues}. The lack of reusable components in the design of these software implementations motivates the design and implementation of our library in this work.

\subsection{Libraries for Out-of-core Implementations of Accelerator Kernels}

The CUBLAS-XT library \cite{CUBLAS-XT} provides a set of BLAS routines that utilize multiple GPUs connected to the same motherboard. It uses CUDA streams \cite{cuda-streams} and events to efficiently manage data transfers across PCI-Express bus and kernel invocations on the GPUs. The routines in the library also support out-of-core operation where the size of the matrices are limited only by the system memory size.

MAGMA \cite{MAGMA} provides out-of-core dense LU, Cholesky, and QR factorizations.

Victream \cite{Suzuki2017} is a directed acyclic graph (DAG) computing framework for out-of-core computations on multiple GPUs. At the heart of Victream is a scheduler that employs locality-aware scheduling and data prefetching for performance optimization.

\subsection{Stream Libraries Supporting Accelerator Kernel Computations}

CUDA streams \cite{cuda-streams} facilitate efficient overlapping of data transfers (from host to device or device to host) and kernel computations on the device. All device operations (kernel execution, data transfers from host to device or device to host) take place in a stream (``null stream'' or default stream if not specified). Since all operations in non-default streams are non-blocking with respect to the host code, to synchronize the host code with operations in a stream, CUDA events are used.

hStreams \cite{hstreams} provides a streaming, task queue abstraction for heterogeneous platforms similar to CUDA streams \cite{cuda-streams} and OpenCL \cite{opencl-commandqueues}.

Our library, libhclooc, presents an uniform interface and implementation for different types of streaming interfaces, CUDA streams \cite{cuda-streams}, PHI offload streams \cite{offload-streams} and OpenCL \cite{opencl-commandqueues}. We intend to integrate hStreams in our library in our future work.

From the survey, we can conclude that there is an abysmal lack of libraries providing an uniform interface for solving efficiently large problem sizes on hybrid platforms containing two ore more state-of-the-art accelerators. This can be a severe shortcoming for effective use of high performance accelerators in the fields of HPC and Big Data. We address this shortage in this work.

\section{libhclooc: Overview} \label{libhcloocoverview}

\begin{figure*}
    \centering
    \includegraphics[width=\textwidth,height=0.3\textwidth]{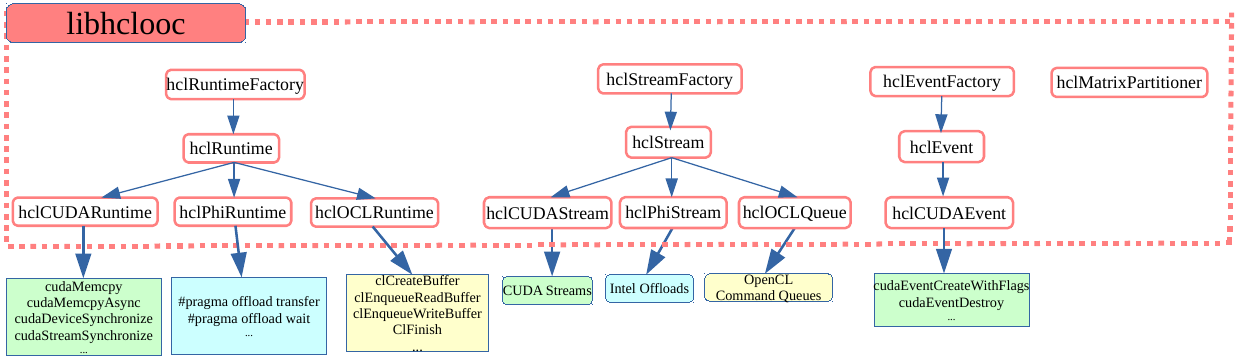}
    \caption{Software organization of libhclooc.}
    \label{fig:softwareorganization}
\end{figure*}

In this section, an overview of libhclooc library is presented. 

The software organization of libhclooc is shown in the Figure \ref{fig:softwareorganization}. The class \emph{hclStreamFactory} contains a factory method to create an instance of \emph{hclStream} class for an input device. The classes \emph{hclCUDAStream}, \emph{hclPhiStream} and \emph{hclOCLQueue} are wrappers around CUDA streams and events, Intel offload streams, and OpenCL command queues providing a simplified programming model for these interfaces.

The class \emph{hclRuntimeFactory} contains a factory method to create an instance of \emph{hclRuntime} class for an input device. The \emph{hclRuntime} class provides an uniform interface for memory allocation and data transfers between host and device. The device-specific classes \emph{hclCUDARuntime}, \emph{hclPhiRuntime}, \emph{hclOCLRuntime}, interact with device-specific classes for streams and events.

The component, \emph{hclMatrixPartitioner}, provides API for partitioning a matrix into blocks that will fit on the device memory.

The simplified programming model will now be presented, which are the runtime interface, how memory is managed, and asynchronous aspects of the library.

\subsection{Programming Model}

The programming model presented by the uniform interface is very explicit. The idea is that the host and device are separate in the sense that the device and host have their own memory, and the host acts as the controller of the device providing it with explicit commands. The host must explicitly issue a command to the device to allocate memory, to transfer data to and from the device, to free memory, and to execute actions on the data. This model allows programmers to more easily reason about the behavior and state of the accelerator.

The API available to the host is provided in the class \texttt{hclRuntime}. An instance of this class is created via the \texttt{hclRuntimeFactory} factory method as either an Intel runtime or CUDA runtime or OpenCL runtime. The runtime is not device specific, only device \emph{type} specific, i.e., one runtime object can be used for many devices of the same type. 

Further abstractions are provided for devices with \texttt{hclDevice}, for streams with \texttt{hclStream}, and events with \texttt{hclEvent}. These classes are primarily data containers, only storing information related to their abstraction, and are used as parameters for the functions within \texttt{hclRuntime}.

\subsection{\texttt{hclRuntime} Methods}

The available functions in \texttt{hclRuntime} are given below with a brief description of their functionality. Their parameters not shown for brevity. 

\begin{itemize}
    \item \texttt{hclMalloc}: Allocate memory on a device.
    \item \texttt{hclFree}: Free memory on a device.
    \item \texttt{hclGetMemSize}: Get the amount of available memory on a device.
    \item \texttt{hclMemCpy}: Copy memory synchronously to or from device.
    \item \texttt{hclMemcpyAsync}: Copy memory asynchronously to or from device.
    \item \texttt{hclDeviceSynchronize}: Block program execution until all preceding tasks on the device are complete.
    \item \texttt{hclStreamSynchronize}: Block stream execution until all tasks in the stream have completed.
    \item \texttt{hclWaitEvent}: Block stream execution until an event is marked as complete.
\end{itemize}

\subsection{Memory Management}

The memory management functionality of libhclooc is very straightforward, consisting of \texttt{hclMalloc} and \texttt{hclFree}. This is very similar to CUDA, but with some restrictions.

\texttt{hclMalloc} can only allocate memory for type \texttt{double}. This is due to a restriction with Intel Offloads where void pointers cannot be used to allocate memory, the pointer needs a concrete type. More implementations of \texttt{hclMalloc} for different pointer types can be created; \texttt{double} was used as that was what was needed initially.

Further, \texttt{hclMalloc} takes a double pointer, \texttt{**ptr}, as the parameter for the pointer to the memory allocation. The inner pointer is used as the device memory pointer after \texttt{hclMalloc} is called. It is important that the parameter is always a pointer to a pointer due to how Intel offloads handle heap allocation. It requires a section of memory to be allocated on the host which will be mapped to the section of memory on the device, which is created for each \texttt{hclMalloc} for Intel devices. This is then freed when \texttt{hclFree} is called. Similarly \texttt{hclFree} takes  \texttt{double} pointers, but can also be extended later.

\texttt{hclGetMemSize} is a helper function to get the total amount of free memory on a device. This function aids the programmers to decide when to invoke in-core or out-of-core kernel.

\subsection{Data Transfers}

Data transfers between host and device again follows a similar model to CUDA. \texttt{hclMemcpy} is used to copy data synchronously, while \texttt{hclMemcpyAsync} is used to copy data asynchronously.

Both functions take an enum, \texttt{Direction}, as a parameter which specifies which whether to copy data from host-to-device (H2D) or device-to-host (D2H). 

\subsection{Asynchronous functions, Streams, and Events}

The interface provides some asynchronous functions to facilitate the \emph{Stream Engine} component of the library. They are identified by the \texttt{Async} suffix. Asynchronous methods are required to be associated with a stream, and optionally an event. The stream is essentially a queue of commands to be executed on a device. When an asynchronous method is called, it is added to the stream. Events provide a way to check if a specific function has completed.

Device specific streams are encapsulated in a \texttt{hclStream} class. A \texttt{hclStream} instance is created using the \texttt{hclStreamFactory} factory method by providing the device object as input. A stream is directly associated with a device.

Device specific events are encapsulated in a \texttt{hclEvent} class. An uninitialised \texttt{hclEvent} variable can be passed as a final argument to asynchronous functions, which will create and initialise the event. The \texttt{hclWaitEvent} function can then be used to wait for the given event to complete. By utilising multiple streams and events, kernel executions can be overlapped on the device with data transfers between the host and the device.

The \texttt{hclStreamSynchronize} method of the libhclooc runtime object will hold program execution until all queued functions in a stream are finished. Similarly, the \texttt{hclDeviceSynchronize} method will hold execution until all functions across all streams on a device are completed.

\section{libhclooc: Design and Implementation} \label{libhcloodesign}

This section covers the design and implementation of \texttt{libhclooc}. The goal is to design a uniform interface for key parts of three disparate accelerator interfaces: CUDA, Intel offloads, and OpenCL command queues.

\subsection{Library Design}

The library was designed to be maintainable and extensible, with the interface being familiar to programmers with prior experience writing code for accelerators. It was designed to feel similar to the CUDA C interface, but in the interest of maintainability it was implemented using C++. 

\subsubsection{Static vs Dynamic Polymorphism}

One of the main reasons for using C++ was the availability of polymorphism and inheritance. As the library unifies three disparate interfaces, one interface is declared with multiple possible implementations underneath it.

C++ offers two forms of polymorphism, static and dynamic. Dynamic polymorphism is resolved at runtime while static is resolved at compile time. Dynamic polymorphism was used due to the flexibility of being able to change implementation without requiring recompilation.

The uniform interface is declared in \texttt{hclRuntime} as a pure virtual base class, and therefore not instantiable. This is then implemented in the derived classes \texttt{hclCUDARuntime}, \texttt{hclPhiRuntime}, and \texttt{hclOCLRuntime} for CUDA, Intel offloads, and OpenCL command queues respectively. An instance of these classes can be created by the \texttt{hclRuntimeFactory} factory method by simply specifying a tuple, a name and an id. This simplistic inheritance structure is easy to understand, provides runtime flexibility, and is simple to extend the interface if needed.

Using dynamic polymorphism will have a performance impact since to determine what implementation to use, a lookup to the virtual function table is required. Static polymorphism will not have this performance impact as implementation is determined at compile time, but the implementation would be more complicated. A more complicated implementation could lead to difficulty maintaining the code, which was decidedly not worth the slight performance improvement. The major bottleneck is the computation being performed on the large problem sizes, which can be many orders of magnitude longer than the time needed to resolve the function implementation. Further, if the specific program did not require the flexibility of changing implementations at runtime, the virtual table lookup may be bypassed by specifying the fully qualified class name for each method call.

\subsection{Implementation}

The device specific interfaces, CUDA, Intel offloads, and OpenCL command queues, are disparate, with different models and behavior. A deep technical understanding of the behavior of underlying interfaces was required to develop a single unified interface. To produce the model presented earlier in this paper for these interfaces involved some implementation compromises, especially with memory management. 

\subsubsection{Memory Allocation}

In CUDA, the host and device memory are clearly separate with explicit functions to allocate memory on the device and copy memory over to an explicit location on the device. Intel attempts to be more transparent, the idea being the memory allocation existing on the host is used as a reference to the related memory on the device. It is possible to implement the appearance of separate memory which must be explicitly managed with Intel offloads but requires careful consideration.

To implement the model of separate host and device memory involved hiding the fact Intel offloads require an allocation on the host of equal size to the equivalent memory allocation on the card. When \texttt{hclMalloc} is called and the Intel Offload implementation is used, a block of memory the same size as the required memory block on the device is created. In practice this can mean there is double memory allocation on the host for any block of memory the programmer wants to copy to the device. The memory block is only allocated in the virtual memory, no data is assigned to the memory so it is not physically allocated. The memory allocation purely acts as a block of address space to map to memory on the device. The doubling up of memory allocation can lead to memory issues as in theory virtual memory will fill up twice as fast, but the maximum virtual address space is exceedingly large (theoretically 16 Exabytes on a 64-bit system) it should not be an issue.

\subsubsection{Intel Offload Pragma Limitations}

Of the two interfaces, Intel offload pragmas is by far the more limited. Pragmas can not be used to allocate memory of a generic type on the device, there must be a concrete type. Instead of the programmer specifying the size of the allocation in bytes, a specific pointer type must be provided along with the length of the memory allocation. The size of the allocation is equal to the length of the allocation multiplied by the byte size of the pointer type. A \texttt{void} pointer type used anywhere in an offload pragma will throw a compile time error. This places restrictions on how generic functions can be. The most common type of variables in scientific kernels are \texttt{doubles}, which were therefore used when required instead of generic pointers. Functions could be made more generic using templates, but function templates are not compatible with abstract classes. Although CUDA can allocate arbitrary number of bytes to a \texttt{void} pointer, it had to conform to the limitations of Intel offloads.

There are also some limitations in relation to events and streams. In CUDA, events are added to a stream; the event is not directly linked with a command, it is marked as complete when the stream execution reaches the event. The Intel offload equivalent of a CUDA event, called signals, must be directly associated with a specific offload. Intel events do not require to be linked to a stream, whereas CUDA events do, which places a restriction on Intel offloads. The result of this is events must be directly related to specific asynchronous functions in the unified interface, which require a stream. This is not necessarily a big limitation for the libraries intended use case, in fact directly associating events with functions can be easier to reason about. 

\section{Out-of-core Matrix-Matrix Multiplication (MMOOC) using libhclooc API} \label{libhcloocmxm}

\begin{figure}[!t]
\small
\lstset{language=C++}
\begin{lstlisting}
hcld *d = hclDeviceFactory::create(name,id);; 
hclRuntime *r = hclRuntimeFactory::create(d);
hclStream **s = hclStreamFactory::create(d,2);
hclMatrixPartitioner(M,N,K,dMemSize,&h,&w,...);
hclEvent *eA[h * w], *rA[h * w], 
*rB[w], *eC[h * w],  *rC[h * w];
for (j = 0; j < w; j++) {
  for (i = 0; i < h; i++) {
    idx=i+j*h; idx1=idx%2; 
    idx2=(idx+1)%2; idx3=idx+1;idx4=idx-1;
    i_=idx3%h;j_=idx3/h;
    if (idx == 0) {
      r->hclMemcpyAsync(d,H2D,s[idx1],&rB[j]);
      r->hclMemcpyAsync(d,H2D,s[idx1],&rA[idx]);
      if ((*step) == 0)
        r->hclMemcpyAsync(d,H2D,s[idx1],&rC[idx]);
    }
    if (idx < (h * w - 1)) {
      r->hclWaitEvent(d,rB[j],s[idx1]);
      r->hclWaitEvent(d,rA[idx],s[idx1]);
      r->hclWaitEvent(d,rC[idx],s[idx1]);
      r->hclDgemmAsync(d,...,s[idx1],&eA[idx]);
      if (idx > 0)
         r->hclWaitEvent(d,eA[idx4],s[idx2]);
      r->hclMemcpyAsync(d,H2D,s[idx2],&rA[idx3]);
      if ((*step) == 0) {
        if (idx > 0)
           r->hclWaitEvent(d,eC[idx4],s[idx2]);
       r->hclMemcpyAsync(d,H2D,s[idx2],&rC[idx3]);
      } else {
        if (idx > 0) {
          r->hclWaitEvent(d,eC[idx4],s[idx2]);
          r->hclMemcpyAsync(d,H2D,s[idx2],&rC[idx3]);
        }
      }
      if (i == (h - 1)) {
        r->hclWaitEvent(d,eA[idx],s[idx2]);
        r->hclWaitEvent(d,eA[idx4],s[idx2]);
        r->hclMemcpyAsync(d,H2D,s[idx2],&rB[j_]);
      }
      if (((*step) == nsteps) || (idx < (h*w-2)))
        r->hclMemcpyAsync(d,D2H,s[idx1],&eC[idx]);
    } else {
      r->hclWaitEvent(d,rB[j],s[idx1]);
      r->hclWaitEvent(d,rA[idx],s[idx1]);
      r->hclWaitEvent(d,rC[idx],s[idx1]);
      r->hclDgemmAsync(d,...,s[idx1], &eA[idx]);
      if ((*step) == nsteps) {
       r->hclMemcpyAsync(d,D2H,s[idx1],&rC[idx3]);
       r->hclWaitEvent(d,rC[idx3],s[idx1]);
      }
    }
  }
}
r->hclStreamSynchronize(d, s[0]);
r->hclStreamSynchronize(d, s[1]);
(*step)++;
\end{lstlisting}
\caption{Out-of-core matrix-matrix multiplication of two matrices A and B of dimensions $M \times K$ and $K \times N$ respectively using libhclooc API.}
\label{fig:libhcloocmxm}
\end{figure}

Figure \ref{fig:libhcloocmxm} shows the out-of-core implementation of matrix-matrix multiplication using libhclooc API. The implementation computes $C = \alpha \times A B + \beta \times C$, where $A$, $B$, and $C$ are matrices of dimensions $M \times K$, $K \times N$, and $M \times N$, respectively and $\alpha$ and $\beta$ are constant floating-point numbers.

The inputs to the implementation are the matrices $A$, $B$, $C$, $\alpha$, $\beta$, the tuple representing the device $\{name,id\}$, the memory size of the accelerator $dMemSize$, and the number of invocations of the implementation given by $nsteps$. The parameter $nsteps$ represents the fact that this out-of-core implementation could be called as a subroutine, for example in parallel matrix-matrix multiplication such as Scalable Universal Matrix Multiplication  Algorithm (SUMMA) \cite{summa}, Hierarchical SUMMA (HSUMMA) \cite{hsumma}, which contain number of main steps equal to $nsteps$.

The implementation is executed on a device represented by a tuple, $\{name,id\}$. For a device that is GPU with ID 0, the device is represented by the tuple, \{``GPU'',0\}. A Xeon Phi device with ID 0 is represented by a tuple, \{``PHI'',0\}. Similarly, a FPGA device with ID 0, the device is represented by the tuple, \{``FPGA'',0\}. 

The handle to \emph{libhclooc} runtime is created in Line 2 using the device object as input. Two streams are then created in Line 3 using the stream factory method. Since creating a new stream has some overhead, we exploit and reuse just two streams in a round robin order so that while one stream is involved in doing computation, the other is transferring data across the PCI-E link.

The function \emph{hclMatrixPartitioner} splits matrix $A$ into $h$ equal horizontal slices, matrix $B$ into $v$ equal vertical slices, and matrix $C$ into $h \times v$ equal rectangular blocks ensuring that the data required for updating of any two blocks of $C$ in the same column is small enough to fit in the accelerator's memory given by the size $dMemSize$.

To synchronize the computations on the device and the transfers of slices of $A$ and $B$ and blocks of $C$, five sets of events are created (Line 5-6).

The implementation consists of $w \times h$ main steps. In each main step, a $c_{ij}$ block is computed. Lines 12-17 contains data transfers of slices $dA[0]$, $dB[0]$, and block $dC[0]$ using stream 0 from host to device (represented by the macro H2D). Then, three events $rA[j]$, $rB[idx]$, and $rC[idx]$ are recorded in the stream (Lines 19-21). The process then waits for the events to be signaled, which takes place after the completion of the data transfers. When the events are signaled, in-core DGEMM kernel invocation (Line 22) is invoked on the sub-matrices ($dA[i]$, $dB[j]$, and $dC[idx]$). 

\begin{figure}[!t]
	\centering
	\includegraphics[width=3in]{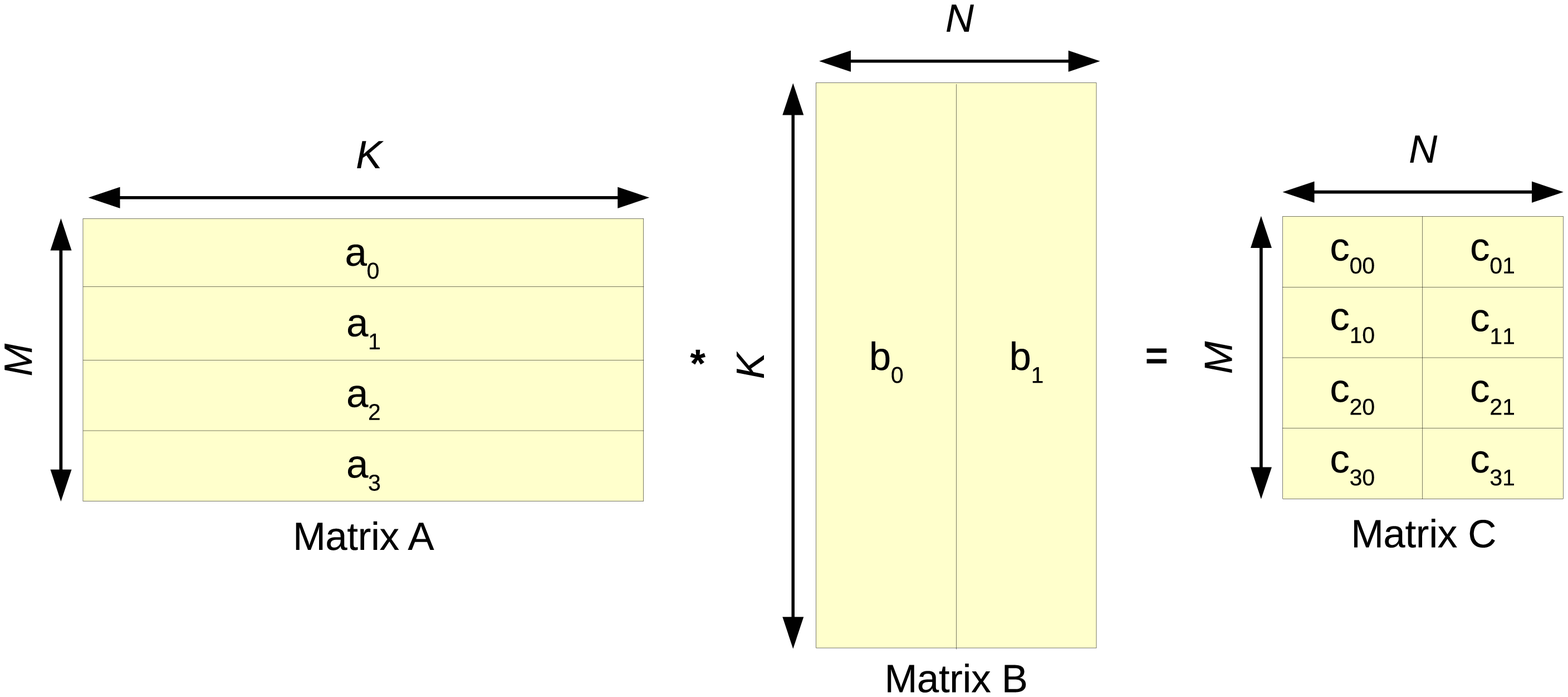}
	\caption{Decomposition of matrix $A$ into 4 horizontal slices, matrix $B$ into 2 vertical slices, and matrix $C$ into 8 ($= 4 \times 2$) blocks.}
	\label{fig:matrix_split}
\end{figure}

\begin{figure*}[!t]
	\centering
	\includegraphics[width=7in,height=2.5in]{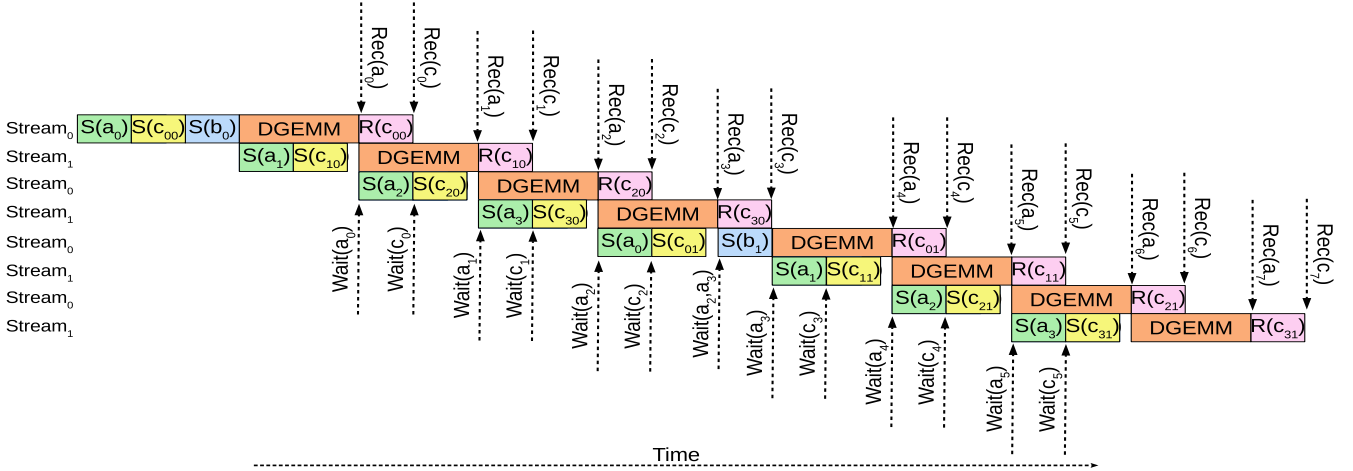}
	\caption{Pipeline structure for the out-of-core matrix-matrix implementation of sample matrices shown in Figure \ref{fig:matrix_split}. Concurrent data transfers in two directions are represented by $S()$ calls and overlapping of data transfers and kernel executions (represented as DGEMM). Events, $Rec(x)$ and $Wait(x)$, are used for synchronization of data transfers.}
	\label{fig:pipelineStructure}
\end{figure*}

To illustrate the software pipeline employed in the out-of-core implementation, we consider a simple example containing three matrices $A$, $B$ and $C$ whose dimensions are $4 \times 8, 8 \times 4$, and $4 \times 4$ respectively. We will assume the main memory size of the accelerator to be 44 elements. Since the total workload size (elements) of the matrices (80) exceeds 44, the matrices are partitioned as shown in the Figure \ref{fig:matrix_split}. 

The software pipeline is composed of five stages described below:
\begin{itemize}
	\item \textbf{S($b_i$):} Sending the i\_th slice of matrix $B$ ($b_i$) from host to device.
	\item \textbf{S($a_i$):} Sending the i\_th slice of matrix $A$ ($a_i$) from host to device.
	\item \textbf{S($c_{ij}$):} Sending a rectangular block of matrix $C$ ($c_{ij}$) from host to device.
	\item \textbf{DGEMM:} Vendor-supplied optimized in-core DGEMM invocation computing the matrix product $c_{ij} = \alpha \times a_i b_i + \beta \times c_{ij}$.
	\item \textbf{R($c_{ij}$):} Sending the updated block $c_{ij}$ of $C$ back from device to host.
\end{itemize}

To make sure data stored in device buffers is not overwritten until kernel executions that operate on the data have completed, events are created for each sub-matrix in $A$ and $C$. As shown in the figure \ref{fig:pipelineStructure}, $Rec(x)$ represents recording the event associated with block $x$, and $Wait(x)$ makes the process wait for the event associated with block $x$ until it is recorded.

This software pipeline uses two streams and therefore performs very efficiently on Nvidia GPUs, which provide separate copy engines for data transfers between host and device, device and host, and engines for kernel invocation.

One can see that there is lot of synchronization code using events to overlap computations and communications correctly and effectively. However, this synchronization pattern is common and can be reused for out-of-core implementations of other data-parallel kernels. In our future work, we will consider design and development of a pattern language (a domain specific language (DSL)) that programmers can use to specify the software pipeline at higher level of abstraction. The compiler for this language will then generate all the boilerplate synchronization code thus removing this burden from the programmers.

\section{Experimental Results} \label{experimentalresults}

\begin{table}
\caption{HCLServer1: Specifications of the Intel Haswell multicore CPU, Nvidia K40c, and Intel Xeon Phi 3120P.}
\label{table:hclserver1}
\centering
\begin{tabular}{ |l|l| }
  \hline
  \multicolumn{2}{|c|}{\textbf{Intel Haswell E5-2670V3}} \\ \hline
	No. of cores per socket & 12 \\ \hline
	Socket(s) & 2 \\ \hline
	CPU MHz & 1200.402 \\ \hline
    L1d cache, L1i cache  & 32 KB, 32 KB \\ \hline
    L2 cache, L3 cache & 256 KB, 30720 KB \\ \hline
    Total main memory & 64 GB DDR4 \\ \hline
	Memory bandwidth & 68 GB/sec \\ \hline
  \multicolumn{2}{|c|}{\textbf{NVIDIA K40c}} \\ \hline
	No. of processor cores & 2880 \\ \hline
	Total board memory & 12 GB GDDR5 \\ \hline
	L2 cache size & 1536 KB \\ \hline
	Memory bandwidth & 288 GB/sec \\ \hline
  \multicolumn{2}{|c|}{\textbf{Intel Xeon Phi 3120P}} \\ \hline
	No. of processor cores & 57 \\ \hline
	Total main memory & 6 GB GDDR5 \\ \hline
	Memory bandwidth & 240 GB/sec \\ \hline
\end{tabular}
\end{table}

We perform our experiments on two research servers, \emph{HCLServer1} and \emph{HCLServer2}. \emph{HCLServer1} contains an Intel Haswell multicore CPU, Nvidia K40c GPU, and Intel Xeon Phi 3120P whose specifications are given in the Table \ref{table:hclserver1} respectively. The FPGA in our server is the AlphaData ADM-PCIE-7V3 accelerator card. This device features a Xilinx Virtex 7 690T FPGA, with 16GB of DDR3 DRAM at 1333MT/s. The FPGA devices operates at a max TDP of 25W in sharp contrast to the other accelerators such as GPU and Xeon Phi with TDPs of 235W and 300W respectively. The OS on it is CentOS 7.2.1511.

\begin{table}
\caption{HCLServer2: Specifications of the Intel Skylake multicore CPU and Nvidia P100 PCIe.}
\label{table:hclserver2}
\centering
\begin{tabular}{ |l|l| }
  \hline
  \multicolumn{2}{|c|}{\textbf{Intel Xeon Gold 6152}} \\ \hline
	Socket(s) & 1 \\ \hline
	Cores per socket & 22 \\ \hline
       L1d cache, L1i cache  & 32 KB, 32 KB \\ \hline
       L2 cache, L3 cache & 256 KB, 30976 KB \\ \hline        
	Main memory &  96 GB \\ \hline
  \multicolumn{2}{|c|}{\textbf{NVIDIA P100 PCIe}} \\ \hline
	No. of processor cores & 3584 \\ \hline
	Total board memory & 12 GB CoWoS HBM2 \\ \hline
	Memory bandwidth & 549 GB/sec \\ \hline
\end{tabular}
\end{table}

\emph{HCLServer2} contains an Intel Skylake multicore CPU and Nvidia P100 PCIe GPU whose specifications are given in the Table \ref{table:hclserver2} respectively. The OS on it is Ubuntu 16.04 LTS.

To obtain an experimental data point, the application is executed repeatedly until the sample mean lies in the 95\% confidence interval and a precision of 0.025 (2.5\%) has been achieved. For this purpose, Student's t-test is used assuming that the individual observations are independent and their population follows the normal distribution. We verify the validity of these assumptions using Pearson's chi-squared test. When we mention a single number such as floating-point performance (in TFLOPs), it is assumed that we are referring to the sample mean determined using the Student's t-test.

We perform two sets of experiments. In the first set, we compare the performance of our MMOOC implementation written using libhclooc API (Figure \ref{fig:libhcloocmxm}) with the state-of-the-art accelerator-specific implementations \cite{Khaleghzadeh2018} (ZZGemmOOC for Nvidia CUDA \cite{ZZGemmOOC} and XeonPhiOOC for Intel Xeon Phi \cite{XeonPhiOOC}) on HCLServer1. We do not report any results for libhclooc executing MMOOC on the FPGA since the basic in-card OpenCL matrix-matrix multiplication implementation is very poor.

In the second set, we compare the performance of our MMOOC implementation with the state-of-the-art accelerator-specific implementation (ZZGemmOOC for Nvidia CUDA \cite{ZZGemmOOC}) and CUBLAS-XT \cite{CUBLAS-XT} using the latest generation Nvidia P100 PCIe GPU on HCLServer2.

The performance (execution speed) of matrix multiplication for two dense matrices of sizes $M \times K$ and $K \times N$ is calculated as $\frac{2 \times M \times K \times N}{t}$, where \emph{t} is the execution time including the data transfers between host and device. The range of problem sizes ($N$) tested is $\{1024,2048,\cdots,46080\}$. For Nvidia GPUs, libhclooc switches to out-of-core operation when $N$ exceeds 22528. For Intel Xeon Phi, libhclooc switches to out-of-core operation when $N$ exceeds 16384.

\begin{figure}[!t]
\centering
\subfloat[][]{
\includegraphics[width=3.2in, height=1.7in]{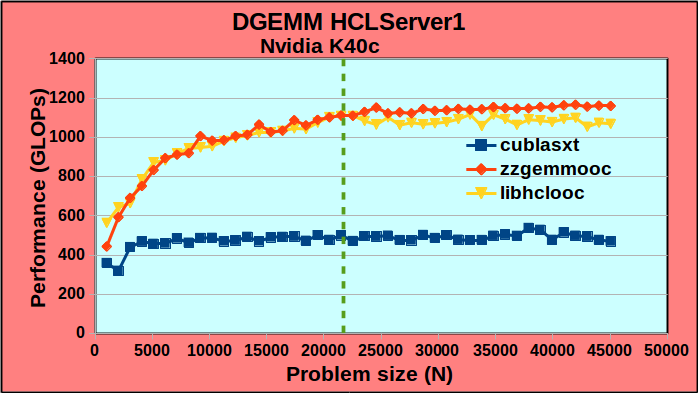}
\label{fig:libhcloochclserver1gpu}}\hfill
\subfloat[][]{
\includegraphics[width=3.2in, height=1.7in]{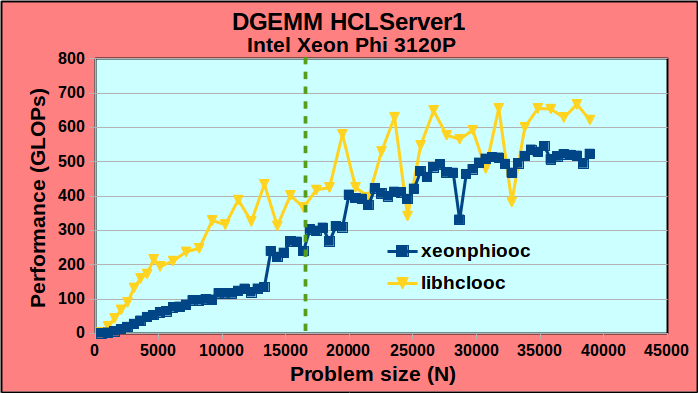}
\label{fig:libhcloochclserver1phi}}\hfill
\subfloat[][]{
\includegraphics[width=3.2in, height=1.7in]{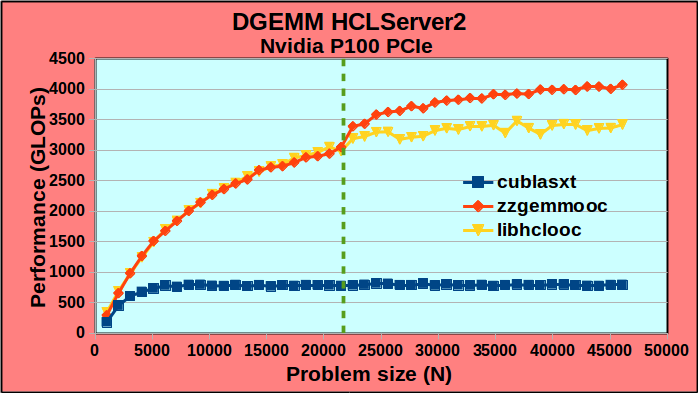}
\label{fig:libhcloochclserver2gpu}}
\caption{a). Comparison of performances of CUBLAS-XT, ZZGemmOOC, and libhclooc for MMOOC on Nvidia K40c on HCLServer1. b). Comparison of performances of XeonPhiOOC and libhclooc for MMOOC on Intel Xeon Phi 3120P on HCLServer1. c). Comparison of performances of CUBLAS-XT, ZZGemmOOC, and libhclooc for MMOOC on Nvidia P100 PCIe on HCLServer2. Green line shows the transition point from in-core to out-of-core execution.}
\end{figure}

Figure \ref{fig:libhcloochclserver1gpu} shows the comparison of performances of MMOOC executed using the Nvidia K40c GPU on HCLServer1. The peak double-precision floating point performance of Nvidia K40c is 1.43 TFLOPs. The ZZGemmOOC implementation reached 1.16 TFLOPs at its peak. The peak double-precision floating point performance of libhclooc was 1.11 TFLOPs. The results therefore show libhclooc performing at 96\% of the peak performance of the ZZGemmOOC implementation. There is also a 0\% loss in performance when the program transitions to out-of-core execution. Both ZZGemmOOC and libhclooc implementations outperform the Nvidia's CUBLAS-XT implementation (by more than 2.3x).

Figure \ref{fig:libhcloochclserver1phi} shows the comparison of performances of MMOOC performed on the Intel Xeon Phi 3120P on HCLServer1. The peak double-precision floating point performance of libhclooc is 667 GFLOPs compared to 549 GFLOPs of the XeonPhiOOC implementation. One can see that libhclooc outperforms the XeonPhiOOC implementation.

It can be assumed that the majority of this loss in performance is due to the abstraction overhead. At least part of this overhead is due to the use of dynamic polymorphism within the library.

The erratic results of the Intel Xeon Phi could be attributed to a few things. When creating an offload stream for a Xeon Phi, it is required to set the number of threads this stream will use. In the libhclooc implementation with two streams, the threads are split in half, ideally with no overlap between threads assigned to each stream. The thread assignment behavior is set using environment variables, but the optimal settings are difficult to discern. What is optimal for smaller matrix sizes will cause the program to crash for larger matrix sizes or cause significant performance issues. Our testing revealed that the two stream implementation is simply not optimal for Intel Xeon Phis. A more optimal setup may involve one thread which would allow operations to utilize all the threads, running parallel calculations at full capacity. 

XeonPhiOOC \cite{XeonPhiOOC} now uses an implementation with a single stream instead of two. Its peak double-precision floating point performance of libhclooc is 725 GFLOPs. Therefore, libhclooc achieves 92\% of the peak. When we reimplemented MMOOC using one stream, these performance drops were not observed. The results, therefore, suggest that the disparate capabilities provided for communication and computation overlap by accelerators must be taken into account when designing and developing out-of-core libraries.

The memory on the Intel Xeon Phi also exhibited some unexplained behavior. After one iteration of the out-of-core kernel, the reported available memory would not return to full capacity and sometimes drastically decreases after further iterations. The next time the program is run the memory is back to full availability. This seems to suggest that the reported available memory from the tool (Intel MPSS \cite{IntelMPSS}) is incorrect. The experiments were run with an override that ignores the reported decrease in memory, as otherwise each iteration would have different size partitions due to differences in memory size.

Figure \ref{fig:libhcloochclserver2gpu} shows the comparison of performances performed on the Nvidia P100 GPU on HCLServer2. The peak double-precision floating point performance of Nvidia P100 is 4.7 TFLOPs. The accelerator-specific optimized ZZGemmOOC implementation reached 3.90 TFLOPs at its peak (83\% of peak). The peak double-precision floating point performance of libhclooc is 3.5 TFLOPs. The results therefore show libhclooc performing at 90\% of the peak performance of the ZZGemmOOC implementation. There is also a 0\% loss in performance when the program transitions to out-of-core execution, which is shown by the green line in the graph. Both ZZGemmOOC and libhclooc implementations outperform the Nvidia's CUBLAS-XT implementation (by more than 4x).

Implementing MMOOC using libhclooc API required 75\% less lines of code (LOC). The interface in libhclooc is a lot more concise and straightforward, especially when compared to using Intel offload pragmas, fulfilling the objective of an easier-to-use interface.

\section{Conclusions and Future Work} \label{conclusions}

Hardware accelerators are increasingly seeing utilisation by extreme-scale high performance computing (HPC), cloud, and Big data platforms to facilitate execution of workloads that demand high energy efficiency (high performance and low energy consumption). These accelerators have unique interfaces and programming models, and have limitations which must be addressed to facilitate execution of large workloads.

This paper presents an implementation of libhclooc, providing a uniform interface for CUDA streams and events, Intel offloads, OpenCL command queues and a reference implementation of efficient out-of-core matrix-matrix multiplication. 

The uniform interface presents a more straightforward programming model. It is more concise, less complex, and easier to understand. Its functions are very explicit, allowing easier reasoning about the state of the accelerator. Using the interface allows programmers to write code to the interface and execute on three different types of accelerators, greatly reducing the amount of code required. 75\% less code is written compared to the three state-of-the-art accelrator-specific implementations.

In terms of performance, the current implementation has an impact on peak performance of 10\% for GPUs and 8\% for Intel Xeon Phis when compared to state-of-the-art accelerator-specific implementations. The out-of-core matrix multiplication presented shows a 0\% performance loss for when the workload exceeds the memory capacity of the device, the performance impact comes from the overhead of wrapping multiple interfaces. Our immediate goal is to find ways to reduce this abstraction overhead.

Intel has developed a new library for heterogeneous computing named hStreams \cite{hstreams}. We plan to add support for this new library in libhclooc to replace Intel offloads entirely.

We plan to add full support for level-3 BLAS kernels in our future work. Furthermore, we plan to provide out-of-core factorizations (LU, QR, Cholesky) that use the out-of-core matrix-matrix multiplication (DGEMM) as a fundamental building block.

We will also look at developing extensions of libhclooc for facilitating programming out-of-core implementations of accelerator kernels for multi-GPU platforms. 

The software implementation of SummaGen presented in this paper is located at \cite{libhclooc}.

\section*{Acknowledgment}

This publication has emanated from research conducted with the financial support of Science Foundation Ireland (SFI) under Grant Number 14/IA/2474.

\bibliographystyle{IEEEtran}
\bibliography{IEEEabrv,paper}

\end{document}